\documentstyle[epsfig]{aa}

\newcommand{\be}{\begin{equation}}
\newcommand{\ee}{\end{equation}}

\begin{document}

\title{Are some breaks in GRB afterglows caused by their spectra?}
\author{D.M. Wei\inst{1,2}, and T. Lu\inst{3,4}}
\institute{Purple Mountain Observatory, Chinese Academy of Sciences, Nanjing, China
\and
      National Astronomical Observatories, Chinese Academy of Sciences, China
\and
       Department of Astronomy, Nanjing University, Nanjing, 210093, China
\and
        Laboratory for Cosmic-Ray and High-Energy Astrophysics, Institute of
        High Energy Physics, Chinese Academy of Sciences, Beijing, 100039, China}

\date{Received date/ Accepted date}

\abstract{Sharp breaks have been observed in the afterglow light
curves of several GRBs; this is generally explained by the jet
model. However, there are still some uncertainties {\bf
concerning} this interpretation due to the unclear hydrodynamics
of {\bf jet} sideways expansion. Here we propose an alternative
explanation to these observed breaks. If we assume that the
multiwavelength spectra of GRB afterglows are not made of exact
power law segments but their slope changes smoothly, i.e.
$d\beta/d log\nu<0$, where $\beta$ is the spectral index, we find
that this fact can very nicely explain the afterglow light curves
showing breaks. Therefore we suggest that some breaks in the
afterglow light curves may be caused by their curved spectra. The
main feature of this interpretation is that the break time is
dependent on the observed frequency, while the jet model produces
achromatic breaks in the light curves. In addition, it is very
important to know the position of the characteristic frequency
$\nu_{c}$ in the multiwavelength spectrum at the time of the
break, since it is a further discriminant between our model and
the jet model. We find that although the optical light curves of
seven GRB afterglows can be {\bf well fitted by the model we
propose}, in fact only one of them (i.e. GRB000926) {\bf can be
explained in this framework, since for other ones the
characteristic frequency $\nu_{c}$ is} either above the optical
after the break or below the optical before the break.
} \maketitle

\section{Introduction}
It is widely accepted that the emission from GRB afterglows can be
well described by the fireball model, in which the ejecta from an
underlying explosion expand into the surrounding medium, {\bf
producing a relativistic shock}(e.g. Piran 1999 and references
therein). In the standard picture, the electrons are accelerated
to relativistic energies, with their Lorentz factors described by
a simple power law distribution $N(\gamma_{\rm e})\propto
\gamma_{\rm e}^{-p}$ above the minimum value $\gamma_{\rm m}$.
Besides particle acceleration, the shock is also responsible for
the creation of a strong magnetic field. Under these conditions
the electrons radiate synchrotron {\bf emission; thus, the
afterglow flux is} $f(t,\nu)\propto t^{\alpha}\nu^{\beta}$, where
the temporal index ($\alpha$) and the spectral index ($\beta$) are
related to $p$ and the dynamics of the blast wave (Wijers et al.
1997; Wei \& Lu 1998; Sari et al. 1998; Huang et al. 2000).

The optical light curves of afterglows can generally be described
by {\bf a single power law decay with index} $\alpha \simeq -1.1
\sim -2$. {\bf However, changes in the light decay rate, with a
transition to a steeper power law behavior have been detected in
several GRBs}(GRB990123: Kulkarni et al. 1999, Castro-Tirado et
al. 1999; GRB990510: Harrison et al. 1999, Stanek et al. 1999;
GRB000301C: Rhoads \& Fruchter 2001, Masetti et al. 2000b;
GRB000926: Sagar et al. 2001a, Price et al. 2001, Fynbo et al.
2001; GRB010222: Masetti et al. 2001, Stanek et al. 2001, Cowsik
et al. 2001, Sagar et al. 2001b; GRB991216: Halpern et al. 2000,
Sager et al. 2000; GRB991208: Castro-Tirado et al. 2001;
GRB990705: Masetti et al. 2000a). Such breaks are usually
explained by the jet model. Rhoads (1997; 1999) and Sari et al.
(1999) have pointed out that the lateral expansion of the
relativistic jet will cause a change in the hydrodynamic behavior
and hence a break in the light curve. However, jet evolution and
emission are very complicated processes, for which different
analytic or semi-analytic calculations lead to different
predictions for the sharpness of the jet break, the jet break time
and the duration of the transition. For example, Rhoads (1999)
claimed that jet expansion produces sharp breaks in the light
curves, while some numerical calculations show that breaks occur
smoothly and gradually (Panaitescu \& Meszaros 1999; Moderski et
al. 2000; Kumar \& Panaitescu 2000; Wei \& Lu 2000a,b). In
particular, the light curve of GRB010222 seems difficult to be
explained by the jet model (Masetti et al. 2001; Dai \& Cheng
2001).

Here we propose an alternative {\bf explanation for these}
observed breaks. We assume that the multiwavelength spectra of GRB
afterglows are not made of exact power law segments, but that
their slope changes smoothly, i.e. $d\beta/d log\nu<0$. We find
that this model can nicely explain at least one of the afterglow
light curves showing breaks. In next section we describe our model
and show the effect of the curved spectra on the afterglow light
{\bf curves; after this, some discussions} and conclusions are
given.

\section{The effect of curved spectra on afterglow light curve}

In the fireball model the afterglow emission spectrum is described
by Sari et al. (1998) as a series of 4 different power law
segments, continuously connecting in correspondence of 3
characteristic synchrotron frequencies ($\nu_{\rm a}$, $\nu_{\rm
m}$ and $\nu_{\rm c}$). However, it is likely that these
connections are not as sharp as in the modelization of Sari et al.
(1998); indeed, in some cases a smooth reconnection over the
synchrotron characteristic frequencies listed above has been
hypothesized (e.g. Granot \& Sari 2001). Therefore, a smooth
spectral bending over $\nu_{\rm a}$, $\nu_{\rm m}$ and $\nu_{\rm
c}$, which implies a curved spectrum, can be considered as a
reasonable {\bf assumption. In this hypothesis, the slope} of the
spectra changes as $d\beta/dlog\nu<0$, where the spectral index
$\beta$ is defined as $\beta=dlogF_{\rm \nu}/d\log \nu$.

Now let us consider the standard case, i.e. the one in which the
blast wave is isotropic and adiabatic, the surrounding medium is
homogeneous, and the afterglow emission is mainly produced by
synchrotron radiation from accelerated electrons. Under these
conditions, Sari et al. (1998) {\bf have computed the
characteristic} synchrotron frequencies \be \nu_{\rm m}=5.7\times
10^{11}(\frac{\epsilon_{\rm
B}}{10^{-2}})^{1/2}(\frac{\epsilon_{\rm e}}{0.1})^{2}
E_{52}^{1/2}t_{\rm d}^{-3/2}\;\;\;\;{\rm Hz}  \ee \be\nu_{\rm
c}=2.7\times 10^{15} (\frac{\epsilon_{\rm
B}}{10^{-2}})^{-3/2}E_{52}^{-1/2}n_{1}^{-1}t_{\rm d}^{-1/2}
\;\;\;\; {\rm Hz}  \ee  where $\nu_{\rm m}$ is the emission
frequency corresponding to the minimum electron Lorentz factor
$\gamma_{\rm m}$, $\nu_{\rm c}$ is the cooling frequency, $E_{52}$
is the fireball energy in units of $10^{52}\,erg$, $\epsilon_{\rm
e}$ and $\epsilon_{\rm B}$ are the {\bf energy fractions of
electrons} and magnetic field respectively, $n_{1}$ is the
surrounding medium density in units of $1\,\,{\bf atom}\,\, {\rm
cm^{-3}}$, and $t_{\rm d}$ is the time since burst in units of
$1\,{\rm day}$. {\bf It is obvious that}, for typical parameters,
$\nu_{\rm m}$ is usually far below the optical band few hours
after the GRB, while {\bf $\nu_{\rm c}$ is generally close to the
spectral} ranges in which breaks are observed (i.e. optical and
X-rays), so we assume that the curved spectrum has the form \be
F_{\nu}=2F_{\nu_{\rm c}} [(\frac{\nu}{\nu_{\rm
c}})^{\beta_{1}}+(\frac{\nu}{\nu_{\rm c}})^{\beta_{2}}]^{-1}
\;\;\;\;for \;\;\;\; \nu >\nu_{\rm m} \ee where
$\beta_{2}>\beta_{1}>0$. Thus for $\nu \ll \nu_{\rm c}$,
$F_{\nu}\propto \nu^{-\beta_{1}}$, and for $\nu\gg \nu_{\rm c}$,
$F_{\nu}\propto \nu^{-\beta_{2}}$. Since for typical parameters
$\nu_{\rm m}\ll \nu_{\rm c}$, we have $F_{\nu_{\rm m}}=
2F_{\nu_{\rm c}}(\frac{\nu_{\rm m}}{\nu_{\rm c}})^{-\beta_{1}}$.
In the standard case, the evolution of the bulk Lorentz factor is
$\Gamma\propto t^{-3/8}$, the peak flux $F_{\nu_{\rm m}}\propto
t^{0}$, and $\nu_{\rm m}/\nu_{\rm c} \propto t^{-1}$ (e.g. Piran
1999 and references therein), so we have $F_{\nu_{\rm c}}\propto
t^{-\beta_{1}}$. Then, for a fixed frequency $\nu_{\rm obs}$, the
afterglow light curve is \be F_{\nu_{\rm obs}}\propto
\frac{(\frac{t}{t_{\rm c}})^{-\frac{3}{2}\beta_{1}}}
{1+(\frac{t}{t_{\rm c}})^{\frac{1}{2}(\beta_{2}-\beta_{1})}}
\;\;\;for \;\; t>t_{\rm m} \ee where $t_{\rm m}$ and $t_{\rm c}$
are the time when $\nu_{\rm m}$ and $\nu_{\rm c}$ cross the fixed
frequency $\nu_{\rm obs}$. So we have $F_{\nu_{\rm obs}}\propto
t^{-\frac{3}{2}\beta_{1}}$ for $t\ll t_{\rm c}$, and $F_{\nu_{\rm
obs}}\propto t^{-(\beta_{1}+\frac{1}{2}\beta_{2})}$ for $t\gg
t_{\rm c}$.

The main feature of this interpretation is that the break time is
dependent on the observed frequency, i.e. the break time is larger
for smaller frequencies. In the present model, the break occurs
when the characteristic frequency $\nu_{\rm c}$ crosses the
observed frequency $\nu_{\rm obs}$. Since $\nu_{\rm c} \propto
t^{-1/2}$ in the standard case, the break time $t_{\rm b}\propto
\nu_{\rm obs}^{-2}$. Therefore the ratio of the break time of
X-ray and optical is $t_{\rm bX}/t_{\rm bo}=(\nu_{\rm X}/\nu_{\rm
o})^{-2}\approx 10^{-6}$. On the contrary, for the jet model
(Rhoads 1999; Sari et al. 1999) or the transition from a
relativistic to a non-relativistic regime (Dai \& Lu 1999) the
break time is achromatic, so it is easy to distinguish between
these breaks and the one produced by a smoothly curved spectrum.
In addition, it is very important to know the position of the
characteristic frequency $\nu_{\rm c}$ in the multiwavelength
spectrum at the time of the break, since it is a further
discriminant between our model and the jet model.

Based on the above results, we have fitted the optical light
curves of seven GRB afterglows in which breaks were detected. In
Figs.1 to 7 we see that this model can fit some of the observed
data very well. However, from Fig.3 {\bf we see that deviations of
the GRB000301C light curve from the fit are evident; this is due
to the existence of flux fluctuations in the optical light curve.
These short time scale variations can be} explained by, for
instance, microlensing (Garnavich et al. 2000), or by the
re-energization of the blast wave, or by irregularities of the
interstellar medium (see e.g. Masetti et al. 2000b). {\bf Thus,
these deviations} are not connected with either the jet model or
with ours. If we ignore these fluctuations, the overall fit
appears more satisfactory, with a reduced $\chi^{2}=1.8$. In
addition, {\bf the reduced $\chi^{2}$ value of GRB010222 is high,
which is due to the possible existence of a further break about 20
days after the burst, and also due to the microvariability in the
observed optical data, so is not connected with our model.}

Besides optical light curves, X-ray light curves are also
available for four GRB afterglows (GRB990510: Pian et al. 2000;
GRB000926: Piro et al. 2001; GRB010222: in't Zand et al. 2001;
GRB991216: Halpern et al. 2000), and it is important to compare
{\bf our model with these light curves} to see whether it could
fit the decays in different {\bf energy ranges, keeping} in mind
that this model foresees an energy-dependent break. We have fitted
the observed X-ray data using the parameters which best fit the
optical light curves ({\bf for the X-ray break time, we adopt
$t_{\rm cx}=t_{\rm c}(\nu_{\rm X}/\nu_{\rm o})^{-2}\simeq
10^{-6}t_{\rm c}$}), and in Figs.4-6 we see that our model {\bf
can fit also the} X-ray light curves of GRB010222, GRB000926 and
GRB991216 quite well (although the reduced $\chi^{2}$ value of
GRB000926 is somewhat {\bf high this is due to the last data point
which seems to indicate a second break} in the light curve (see
Piro et al. 2001). If we ignore the last point, then the reduced
$\chi^{2}$ {\bf value of the fit is 0.93). It is instead obvious
(see Fig.2) that} the X-ray light curve of GRB990510 cannot be
fitted by our model.

However, we note that, for GRB990123 (Castro-Tirado et al. 1999),
GRB010222 (Masetti et al. 2001), GRB991216 (Halpern et al. 2000),
and GRB990510 (Pian et al. 2001), $\nu_{\rm c}$ lies well above
the optical window after the light curve break, while for
GRB000301C (Jensen et al. 2001) the frequency $\nu_{\rm c}$ is
below the optical range before the light curve break, so these
GRBs cannot be explained by this model. {\bf Only for GRB000926
(Harrison et al. 2001) $\nu_{\rm c}$ is located} below the optical
range after the light curve break, thus this break can be
explained by our model.

\begin{figure}
\epsfig{file=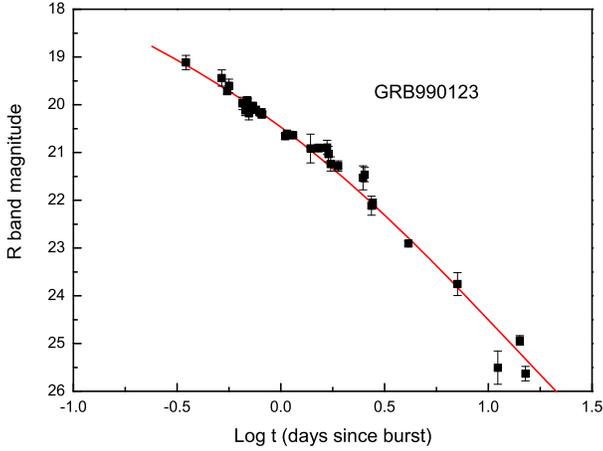, width=8cm} \caption{The R-band light
curve of the GRB990123 afterglow; the contribution of host galaxy
has been subtracted. The solid line represents our best fit
assuming the model described in the text. The fit parameters are:
$\beta_{1}=0.52$, $\beta_{2}=2.8$, $t_{\rm c}=1.2\,{\rm days}$,
the reduced $\chi^{2}=2.1$.}
\end{figure}

\begin{figure}
\epsfig{file=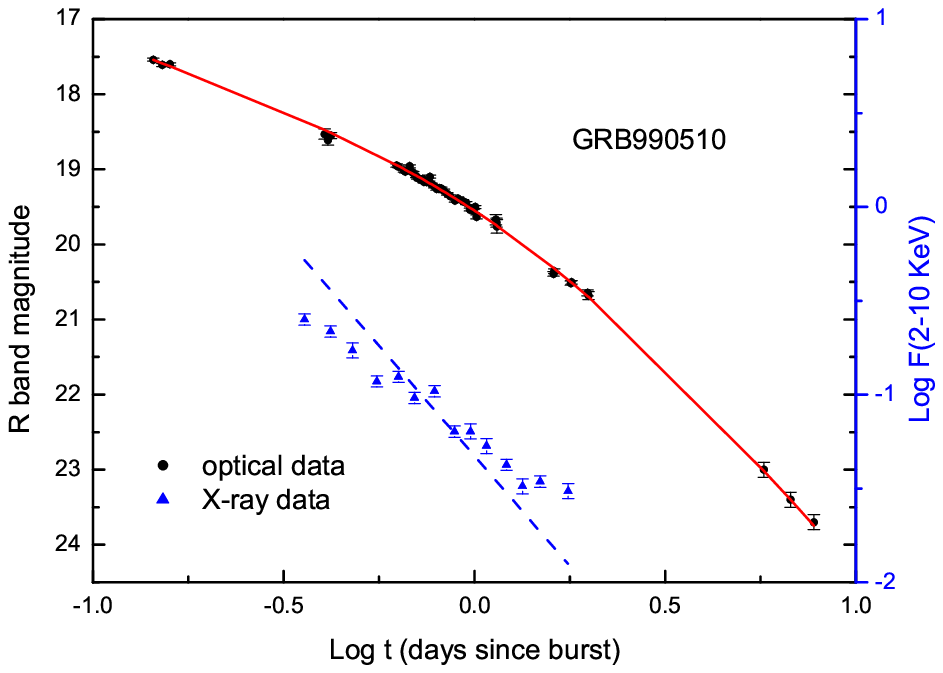, width=8cm} \caption{The R-band and X-ray
light curves of the GRB990510 afterglow. The solid line represents
our best fit to the optical data assuming the model described in
the text. The fit parameters are: $\beta_{1}=0.49$,
$\beta_{2}=3.7$, $t_{\rm c}=1.4\,{\rm days}$, the reduced
$\chi^{2}=2.13$. The dashed line represents the fit to the X-ray
data using the above parameters, the value of reduced $\chi^{2}$
for X-ray fitting is 33.}
\end{figure}

\begin{figure}
\epsfig{file=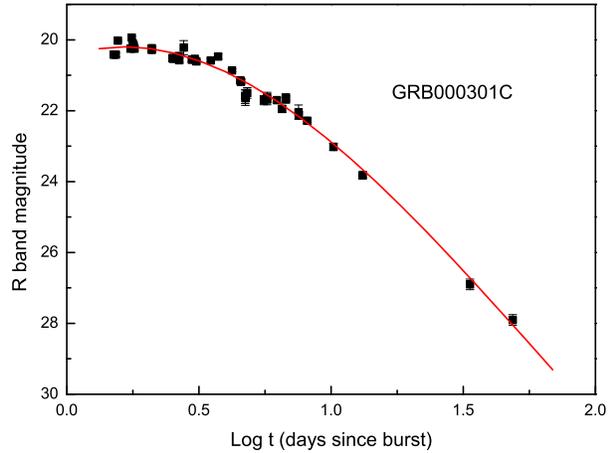, width=8cm} \caption{The R-band light
curve of the GRB000301C afterglow. The solid line represents our
best fit assuming the model described in the text. The fit
parameters are: $\beta_{1}=0.2$, $\beta_{2}=5.6$, $t_{\rm
c}=4.9\,{\rm days}$, the reduced $\chi^{2}=5.8$.}
\end{figure}

\begin{figure}
\epsfig{file=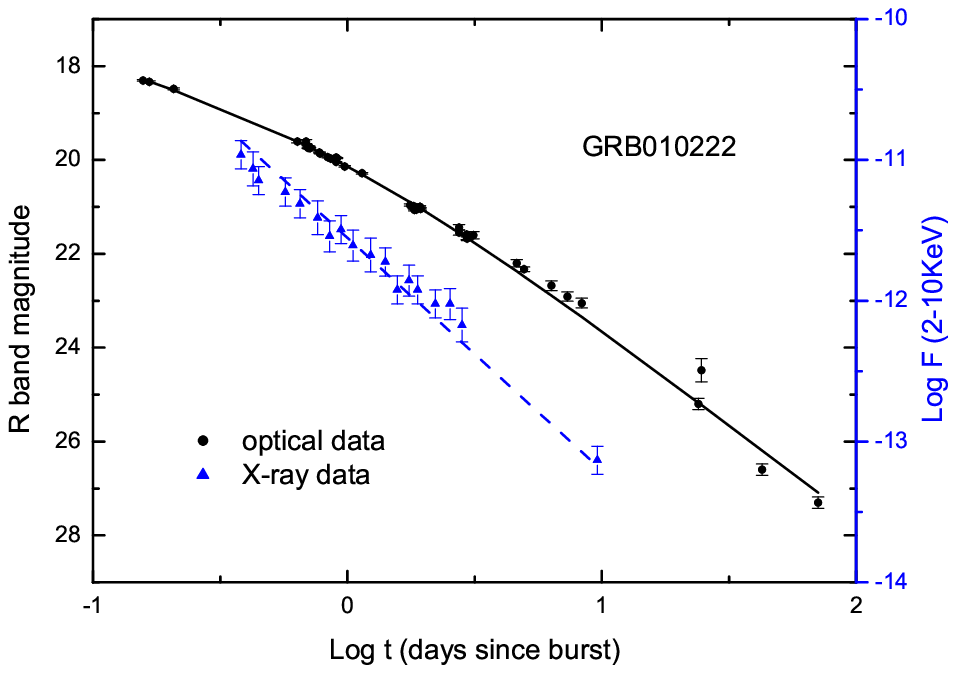, width=8cm} \caption{The R-band and X-ray
light curves of the GRB010222 afterglow. The solid line represents
our best fit to the optical data assuming the model described in
the text. The fit parameters are: $\beta_{1}=0.4$,
$\beta_{2}=2.5$, $t_{\rm c}=0.9\,{\rm days}$, the reduced
$\chi^{2}=6.1$. The dashed line represents the fit to the X-ray
data using the above parameters, the value of reduced $\chi^{2}$
for X-ray fitting is 0.86.}
\end{figure}

\begin{figure}
\epsfig{file=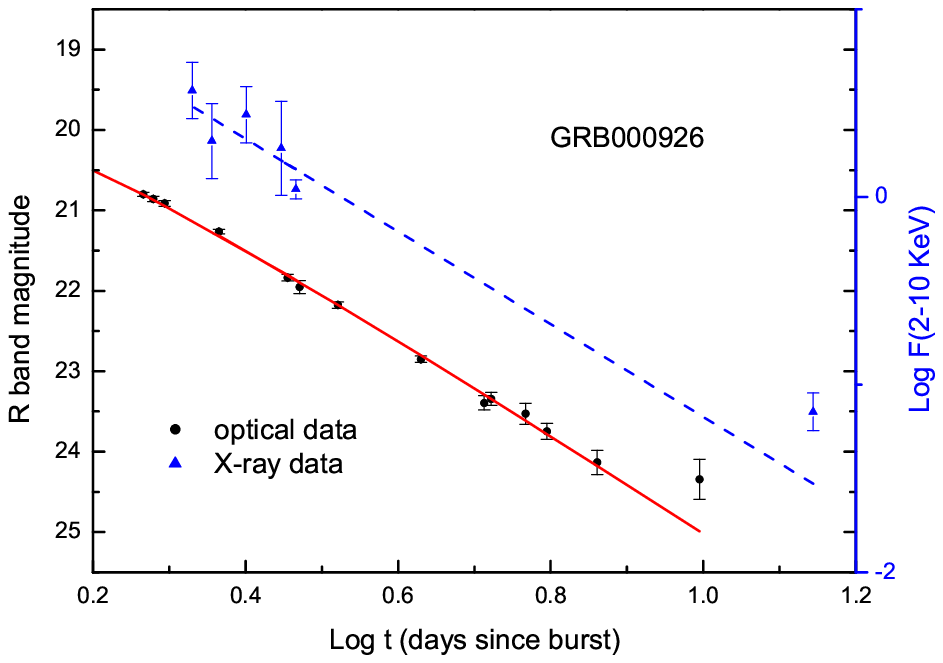, width=8cm} \caption{The R-band and X-ray
light curves of the GRB000926 afterglow. The solid line represents
our best fit to the optical data assuming the model described in
the text. The fit parameters are: $\beta_{1}=0.37$,
$\beta_{2}=4.2$, $t_{\rm c}=1.0\,{\rm days}$, the reduced
$\chi^{2}=1.49$. The dashed line represents the fit to the X-ray
data using the above parameters, the value of reduced $\chi^{2}$
for X-ray fitting is 4.1.}
\end{figure}

\begin{figure}
\epsfig{file=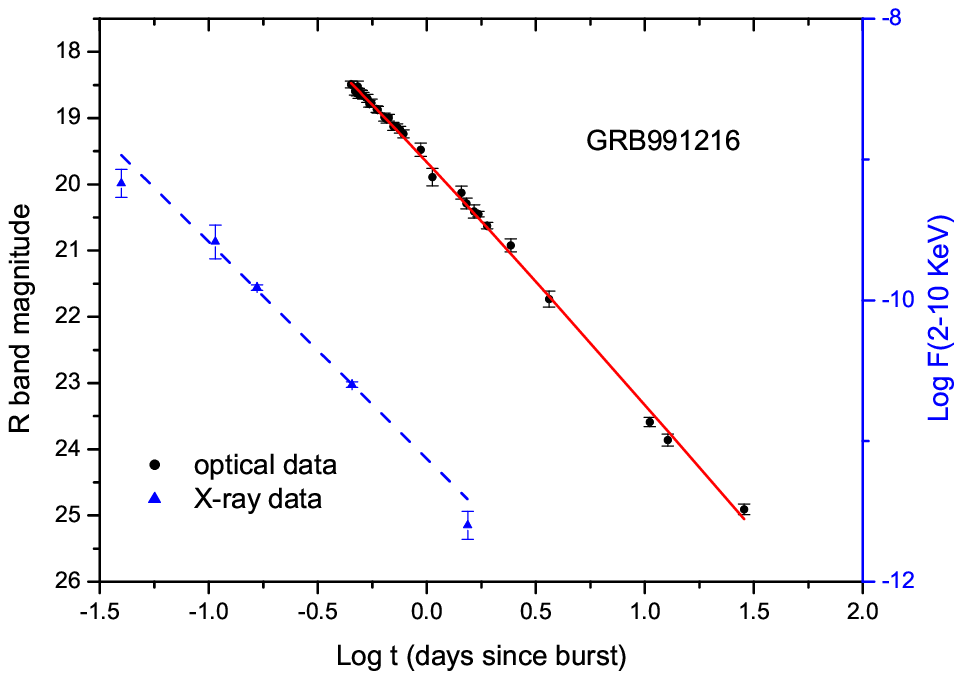, width=8cm} \caption{The R-band and X-ray
light curves of the GRB991216 afterglow. The solid line represents
our best fit to the optical data assuming the model described in
the text. The fit parameters are: $\beta_{1}=0.54$,
$\beta_{2}=2.0$, $t_{\rm c}=0.11\,{\rm days}$, the reduced
$\chi^{2}=0.94$. The dashed line represents the fit to the X-ray
data using the above parameters, the value of reduced $\chi^{2}$
for X-ray fitting is 2.}
\end{figure}

\begin{figure}
\epsfig{file=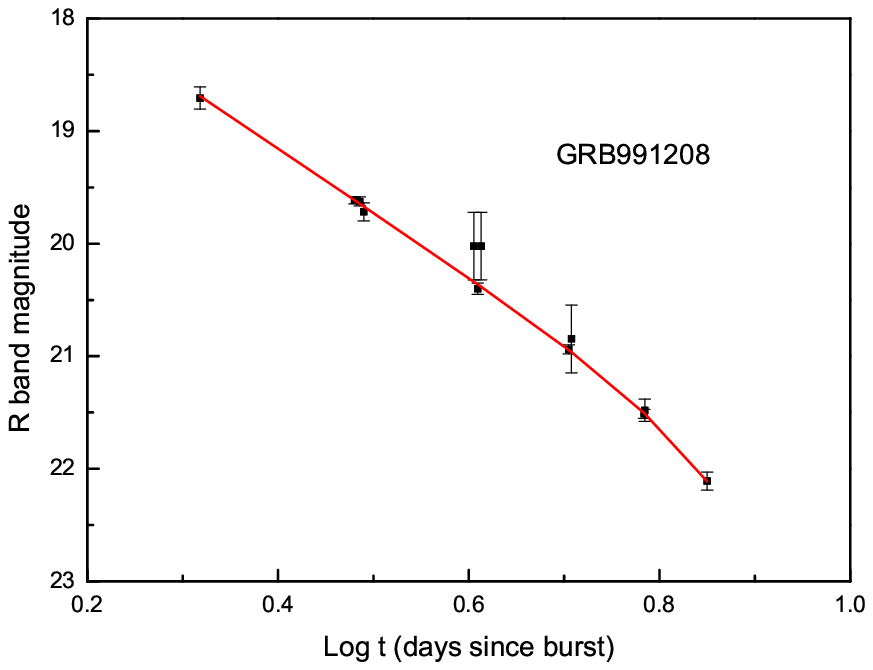, width=8cm} \caption{The R-band light
curve of the GRB991208 afterglow. The solid line represents our
best fit assuming the model described in the text. The fit
parameters are: $\beta_{1}=1.5$, $\beta_{2}=14.6$, $t_{\rm
c}=8\,{\rm days}$, the reduced $\chi^{2}=0.5$.}
\end{figure}

\section{Discussion and conclusions}
In this paper we assumed that GRB afterglow spectra {\bf are not
made of simple power law segments, but} the spectral index changes
gradually with frequency across the afterglow characteristic
synchrotron frequencies. Under these conditions, we have shown
that, even in the standard afterglow model, the afterglow optical
light curves showing breaks can be fitted very well and, except
for GRB990510, the afterglow X-ray light curves can also be fitted
quite well. So we propose that the effects of the curved spectra
on the light curves should not be ignored.

For the simple spectral form we adopted here (Eq.3), considering a
fixed frequency $\nu_{\rm obs}$, the spectral index $\beta$ is
assumed to change with time, so it is of paramount importance to
accurately monitor the GRB afterglow spectral evolution to test
the hypotheses behind the model presented here. In addition, it is
also necessary to improve the energy and time resolution of the
observations in order to verify the presence of a small curvature
of the spectra.

Here we simply take the {\bf curved spectra in the form} of
Eq.(3), but several physical mechanisms may be responsible for
this. {\bf The curved spectra may be caused by intrinsic or
intervening absorption, or by steepening of the electron energy
distribution, or by smooth connection} of the power-law segments
over the characteristic synchrotron frequencies.

In summary, here we have shown that the curved spectra can produce
sharp breaks in the afterglow light curves; however, not all
breaks can be explained by this effect. We think that some of the
steepenings observed in the afterglow light curves may be the
combined result of a curved spectrum and of a collimated fireball.
Future observations will test this hypothesis.

\acknowledgements We thank the referee for several important comments
that improved this paper. This work is supported by the National
Natural Science Foundation (10073022 and 19973003) and the National 973
Project on Fundamental Researches of China (NKBRSF G19990754).

\end{document}